\documentclass[aps,prb,amsmath,amssymb,superscriptaddress,showpacs,twocolumn,floatfix]{revtex4}
\usepackage{amsfonts, relsize}
\usepackage{graphics}
\usepackage{graphicx}
\usepackage{hyperref}

\usepackage{color}
\begin{document}

\title{Competing charge-density-wave, magnetic and topological ground states at and near Dirac points in graphene in axial magnetic fields}

\author{Bitan Roy}
\affiliation{ Condensed Matter Theory Center, Department of Physics, University of Maryland, College Park, MD 20742, USA}

\author{Jay D. Sau}
\affiliation{ Condensed Matter Theory Center, Department of Physics, University of Maryland, College Park, MD 20742, USA}

\date{\today}
\begin{abstract}
In the presence of axial magnetic fields that can be realized in deliberately buckled monolayer graphene, quasi-relativistic Dirac fermions may find themselves in a variety of broken symmetry phases even for weak interactions. Through a detailed Hartree self-consistent numerical calculation in finite strained graphene with cylindrical and open boundaries we establish the possibility of realizing a charge-density-wave order for the spinless fermions in the presence of weak nearest-neighbor repulsion. Such an instability gives rise to a staggered pattern of average fermionic density between bulk and boundary of the system as well as among two sublattices of graphene, due to the spatial separation of the zero energy states localized on opposite sublattices. Although with fermions spin restored, an unconventional magnetic order driven by the onsite repulsion, possibly leads to the dominant instability at the Dirac point, the proposed charge-density-wave order can nevertheless be realized at finite doping, which is always accompanied by a finite ferromagnetic moment. Additionally, the charge-density-wave phase supports a quantized charge or spin Hall conductivity when its formation away from the Dirac point is further preceded by the appearance of topological anomalous or spin Hall insulator respectively. The topological orders in strained graphene can be supported by weak second neighbor repulsion, for example. Therefore, depending on the relative strength of various short-range components of the Coulomb interaction a number of broken symmetry phases can be realized within the zero energy manifold in strained graphene.
\end{abstract}


\maketitle

\vspace{10pt}

\section{Introduction}

Pseudo-relativistic Dirac Hamiltonian frequently emerges as an effective low energy description of quasi-particles in a number of modern day condensed matter systems, such as three dimensional strong $Z_2$ topological insulators and Weyl semimetals. In two dimensions, Dirac fermions can be realized in graphene and surface states of topological insulators. The quasi-particle Fermi velocity in this materials plays the role of the velocity of light and often the real spin of the fermions is replaced by appropriate \emph{pseudo-spin} degrees of freedom. The pseudo-relativistic Dirac equation permits coupling of fermions with \emph{axial gauge fields}, besides the one with regular electro-magnetic potential. The axial magnetic field in three-dimensional strong spin-orbit coupled Dirac materials may result from the {\em Zeeman coupling} \cite{Zhang-modelHamil-TI, Goswami-Roy-axionfieldtheory, chernudub}, whereas that in graphene can arise from its deliberate wrinkling\cite{guinea}. Thus, investigation of the effect of axial magnetic fields in quasi-relativistic systems is of fundamental importance and also experimentally pertinent \cite{pseudofield-experiment-1, pseudofield-experiment-2, artificialgraphene}. In this paper, we focus on its effect in a prototype two dimensional Dirac system, graphene \cite{wallace, semenoff}.

The time-reversal symmetric axial or pseudo magnetic field in graphene, just like the true one, quenches the conical Dirac dispersion into a set of discrete and highly degenerate \emph{fictitious} Landau levels, placed at well separated energies when and only if it is uniform. However, the one at precise zero energy is topologically protected and quite robust against smooth spatial deformation of the strain fields \cite{herbut-pseudocatalysis, aharonov, thesis}. Otherwise, it receives support from only one of the sublattices (say $A$) of honeycomb lattice. On the other hand, the inevitable existence of a \emph{boundary} in a finite strained graphene accommodates an equal number of zero energy states in its surrounding which are localized on the opposite sublattice ($B$). These states are naturally discarded in an infinite system as they grow exponentially towards the boundary \cite{herbut-roy-pseudocat}. In this work we propose a specific modulation of the nearest-neighbor (NN) hopping amplitude in graphene with cylindrical boundary (see Fig. 1) that mimics the effect of axial magnetic field and captures all the above mentioned features of the zero energy subband.

Taking this entire zero energy sub-band into account we show through detail numerical Hartree self-consistent calculations that strained graphene can allow the formation of charge-density-wave (CDW) even when NN repulsion ($V_1$) is weak by keeping only half of the zero energy states localized in one region (either bulk or boundary) or on one sublattice of the system filled. Such ordering, however, does not depend on the detail of system's geometry and we provide numerical evidence of CDW ordering in strained graphene with cylindrical as well as open boundaries for a wide range of \emph{subcritical} ($V_1 < (V_1)_c$) NN interaction, where $(V_1)_c$ is the critical NN repulsion for the CDW ordering in pristine graphene. In addition, we illustrate that the scaling of CDW order with axial magnetic fields ($b$) is qualitatively similar to the one in the presence of real magnetic fields \cite{herbut-roy-analysitc-scaling, roy-herbut-inhomogeneous}. It exhibits a smooth crossover from linear to sublinear variation with $b$ as the strength of weak NN repulsion gradually increases. Finally an almost perfect $\sqrt{b}$ scaling emerges when the NN repulsion is of the order of the critical one in strain-free graphene $(V_1=(V_1)_c)$.

In additional to sublattice, valley and spin degrees of freedom also participate in the low energy dynamics in graphene and provide extra degeneracies to the zero energy manifold, allowing the formation of various other orders as well. For example, spinless fermions in strained graphene can also condense into the topological quantum anomalous Hall insulator (QAHI) phase, representing a chiral scalar, but time-reversal-odd Dirac mass\cite{HJR, haldane}, for weak next-NN (NNN) interaction ($V_2$) \cite{herbut-pseudocatalysis, thesis, ashwin, herbut-roy-pseudocat}. Therefore, depending on mutual strength of $V_1$ and $V_2$, weakly interacting pseudo-relativistic fermions (spinless) in strained graphene can pair into two topologically distinct Dirac vacua. In this work we show that typically when $V_1>V_2$ the CDW is energetically preferred over the QAHI and vice-versa.

When fermion spin is restored, possibly the onsite Hubbard repulsion drives the leading instability near the Dirac point, supporting a magnetic ground state. Recently, through detailed numerical calculation it has been established that the ground state of Hubbard model in strained graphene, termed as {\em global anti-ferromagnet}\cite{roy-assaad-herbut}, through the formation of two ferromagnetic domains with opposite and equal magnetization in bulk and boundary globally leads to only an anti-ferromagnet order of same sign everywhere in the system, while the space-integrated magnetization gets pinned to zero\cite{roy-assaad-herbut}. Nevertheless, when the chemical potential is placed away from the Dirac point, competing CDW, topological quantum anomalous/spin Hall insulators\cite{abanin-pesin} lead to further splittings of the shifted zero energy subband. We argue that when appearance of CDW is preceded by the formation of global-antiferromagnet, the remaining zeroband goes through a full spin polarization. Hence, CDW ordering at finite doping in strained graphene is accompanied by a ferromagnet order. We also show that by continuously tuning the chemical potential away from the charge neutrality point a rich variety of interaction driven broken symmetry phases can be observed in strained graphene.

Rest of the discussion is organized as follows. In the next section, we illustrate a particular way of realizing axial magnetic field in graphene with cylindrical boundary and demonstrate the sublattice entanglement of the zero modes in that system from the computation of local density of states (LDOS). In Sec. III we consider the NN interaction among the fermions (spinless) and through detailed numerical Hartree self-consistent calculations establish the appearance of the CDW order in strained graphene. Section IV is devoted to study the interplay of CDW and QAHI orders within the framework of $V_1-V_2$ model in deformed half-filled honeycomb lattice. Upon restoring spin of the fermions, we address competition among CDW, magnetic, topological ground states in Sec.~V. A summary of our results and discussions on related issues are presented in Sec.~VI.

\section{Axial magnetic field in graphene}        

\begin{figure}[htb]
\includegraphics[width=5.5cm,height=4.5cm]{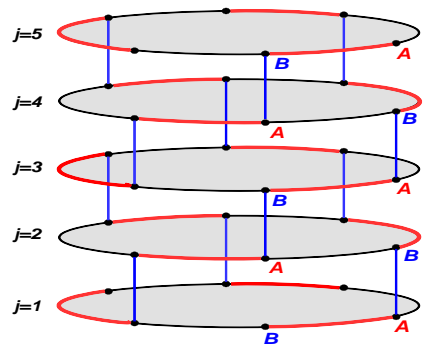}
\caption{(Color online) Proposed modulation of the NN hopping amplitude in honeycomb lattice with cylindrical boundary, yielding a finite axial magnetic field. Red (black) corresponds to stronger(weaker) bond. Along the blue bonds, hopping is unchanged ($t=1$). Red/black bonds experience a gradient along the axis of the cylinder (not shown explicitly). A and B are the two sublattices of honeycomb lattice, and $j$ represents the row index.}
\end{figure}

The axial magnetic field, we consider in this work will be generated through a specific modulation of the NN hopping amplitude, given by\cite{huse} 
\begin{equation}\label{hopaxial}
t_{kl} \rightarrow e^{\chi(k)} \; t \; e^{-\chi(l)},
\end{equation}  
where $k \in A$, $l \in B$. Here $A$ and $B$ represent two sublattices of the honeycomb lattice. The uniform hopping amplitude in strain-free graphene is given by $t$, set to be \emph{unity} hereafter\cite{v1v2units}. To understand the combined effect of the bulk and the boundary, we here consider the cylindrical geometry, shown in Fig.~1\cite{cylinder}. The parameter $\chi$ is chosen such that the hoppings along the vertical bonds in Fig.~1 remain unchanged, while the bonds along the horizontal rows develop an alternating modulation, where the hopping gets stronger or weaker along alternate bonds. Otherwise, the magnitude of the modulation increases monotonically from the bottom to the top end of the cylinder.

With such spatial variation of the hopping amplitude, the effective low-energy Hamiltonian in graphene, dominated by the wave-vectors in the vicinity of $\pm \vec{Q}$ (valleys) \cite{wallace, semenoff}, can be approximated by \cite{herbut-pseudocatalysis, herbut-roy-pseudocat, jckiw-pi, roy-oddQHE} 
\begin{equation}\label{HDpseudo}
H[\vec{a}]=i\gamma_0 \sum_{\nu=1,2} \gamma_\nu \left(\hat{p}_\nu -i \gamma_{3} \gamma_{5} a_\nu (\vec{x}) \right)=e^{\chi(\vec{x}) \gamma_0} H[0] e^{\chi(\vec{x}) \gamma_0}. 
\end{equation}
Here we organize the spinor basis as
\begin{equation}
\Psi^\top=\left[u(\vec{Q}+\vec{p}),v(\vec{Q}+\vec{p}),u(-\vec{Q}+\vec{p}),v(-\vec{Q}+\vec{p}) \right],
\end{equation}
where $|\vec{p}| \ll |\vec{Q}|$. $u$ and $v$ are respectively the fermionic annihilation operators on two inequivalent triangular sublattices of the honeycomb lattice, $A$ and $B$, respectively. Mutually anti-commuting four component $\gamma$ matrices belong to the representation $\gamma_0=\sigma_0 \otimes \sigma_3$, $\gamma_1=\sigma_3 \otimes \sigma_2$, $\gamma_2=\sigma_0 \otimes \sigma_1$, $\gamma_3=\sigma_1 \otimes \sigma_2$, and $\gamma_5=\sigma_2 \otimes \sigma_2$ $\equiv \gamma_0 \gamma_1 \gamma_2 \gamma_3$.\cite{HJR} $\left\{ \sigma_0,\vec{\sigma}\right\}$ are two dimensional identity and the standard Pauli matrices. The first Pauli matrix in each $\gamma$-matrix operates on the valley and the second one on the sublattice index. For simplicity, we here suppress the spin degrees of freedom. The axial vector potential ($\vec{a}$) is related to the parameter $\chi$ according to $a_i(\vec{x})=\epsilon_{ij} \partial_j \chi(\vec{x})$ and the associated axial magnetic field is given by $b(\vec{x})=\partial^2 \chi(\vec{x})$. Thus a particular choice, namely $\chi(j) \propto q j^2$ where $j$ plays the role of the continuum coordinate parallel to the axis of the cylinder produces a {\em roughly uniform} axial magnetic field ($\sim q$) in the system. The axial magnetic field, however, points in opposite direction at two valleys (since $i \gamma_3 \gamma_5=\sigma_3 \otimes \sigma_0$) and thus preserves the time-reversal symmetry. 

\begin{figure}[htb]
\includegraphics[width=4.2cm,height=3.5cm]{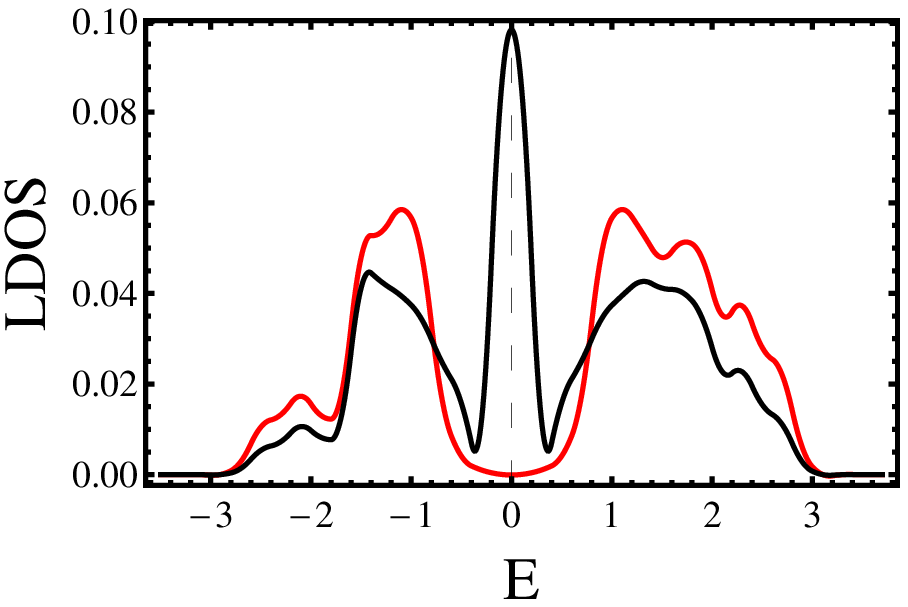}
\includegraphics[width=4.2cm,height=3.5cm]{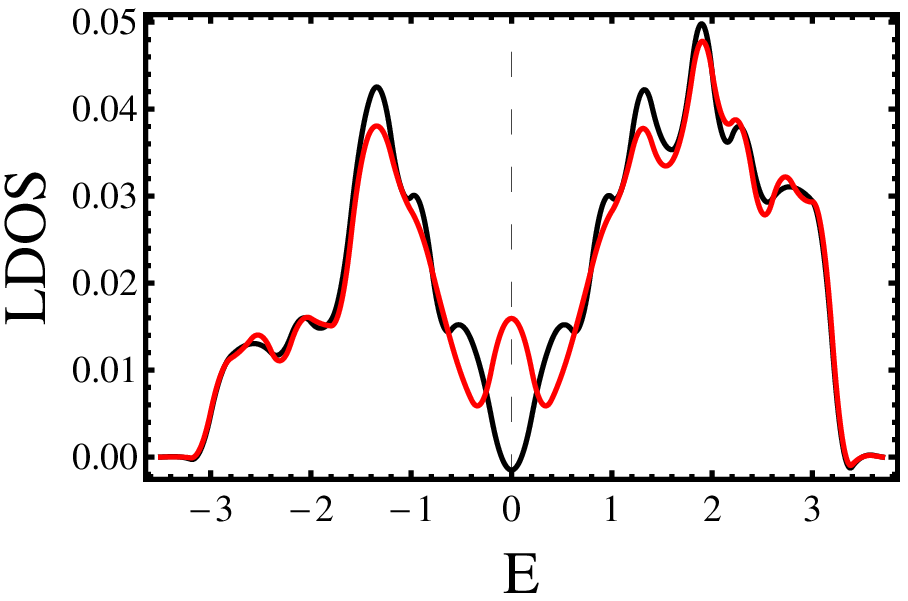}
\includegraphics[width=4.2cm,height=3.5cm]{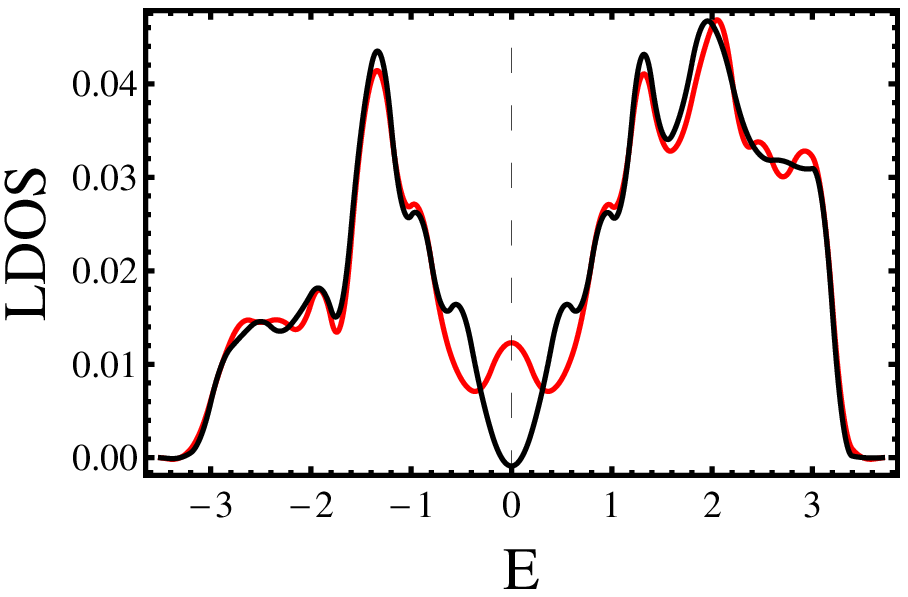}
\includegraphics[width=4.2cm,height=3.5cm]{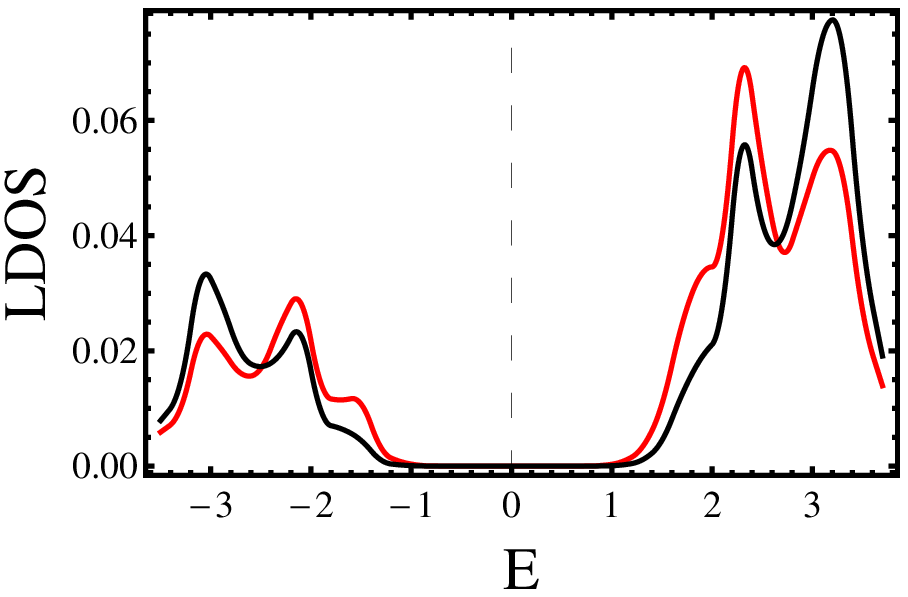}
\caption{(Color online) Plots of LDOS in non-interacting system on sites of A (red) and B(black) sublattices, at the bottom (top, left), near the center (top, right), at the center (bottom, left), and top (bottom, right) row of the cylinder, for $q=0.02$. $E$ is measured in the units of $t(=1)$. We average the density of states over an interval $\Delta E=0.1$, yielding the broadening in LDOS. }
\label{NonIntLDOS}
\end{figure}

Diagonalizing the tight-binding Hamiltonian, with the underlying modulation of hopping, shown in Eq.~(\ref{hopaxial}), in a lattice of $36$(number of sites in each row) $\times \; 25$(total number of rows in the system) sites, a significantly large number of states can immediately be seen in the close vicinity of the zero energy for any $q \geq 0.01$. Here, all the so called zero energy states are slightly shifted away from zero energy due to the finite size effect. We will refer to these low energy states as \emph{zero energy states}. In cylindrical strained graphene, all the zero energy states on the $B$-sublattice are pushed down to the bottom edge, while those residing on the $A$-sublattice are distributed around its center. The top half of the cylinder does not host any zero modes. These features are well resolved in the LDOS on the sites of two sublattices, chosen at various locations of the cylinder, as shown in Fig.~2. The \emph{spikes} in LDOS on A-sublattice at \emph{finite energies} possibly indicate the appearance of the \emph{axial Landau levels}. These findings are in agreement with the continuum description of the problem, in which the zero energy states are\cite{herbut-pseudocatalysis, herbut-roy-pseudocat, jckiw-pi, roy-assaad-herbut}
\begin{equation}
\Psi_{0,n}[\vec{a}](\vec{x})=e^{-\chi(\vec{x}) \gamma_0} \Psi_{0,n}[0](\vec{x}),
\end{equation}
where $n$ represents the total number of zero energy states, which is proportional to the total axial flux quanta enclosed by the system. Since the parameter $\chi$ in our system increases monotonically along the axis of the cylinder and the matrix $\gamma_0$ alters its sign between two sublattices, the normalizable zero energy states live only on A-sublattice, and are localized within the bulk. On the other hand, the amplitude of the zero modes on B-sublattice grows exponentially away from the center of the system. They correspond to non-normalizable zero energy states and appear near the bottom end of the cylinder.

Dirac equation in two spatial dimensions permits coupling of quasi-relativistic fermions with a general time-reversal symmetric $SU(2)$ axial gauge potential, entering the Dirac Hamiltonian as\cite{herbut-pseudocatalysis, roy-oddQHE, FdJuan} 
\begin{equation}\label{genaxial}
H_{gen}[\vec{a}]= i \gamma_0 \sum_{\nu=1,2} \gamma_\nu \left( \hat{p}_\nu - a_\nu^3 \gamma_3 - a_\nu^5 \gamma_5 -a_\nu^{35} i \gamma_3 \gamma_5  \right),
\end{equation}
which can be generated by particular modulation of hopping\cite{herbut-roy-pseudocat} or onsite potential\cite{ ghaemi-ryu-gopalakrishnan} in graphene. However, only the $a_\nu^{35}$ component of the generic $SU(2)$ axial gauge potential preserves the translational symmetry, generated by $i \gamma_3 \gamma_5$\cite{HJR} and for any smooth deformation of the graphene flake, we can safely set $a_\nu^3=a_\nu^5=0$,\cite{roy-oddQHE} whereas $a_\nu^{35} \equiv a_\nu$ in Eq.~(\ref{HDpseudo}). We will discuss the possible routes of realizing $a_\nu^3$ and $a_\nu^5$ components of the axial gauge fields in graphene towards the end of Sec.~IV. 

\begin{figure}[htb]
\includegraphics[width=4.2cm,height=3.5cm]{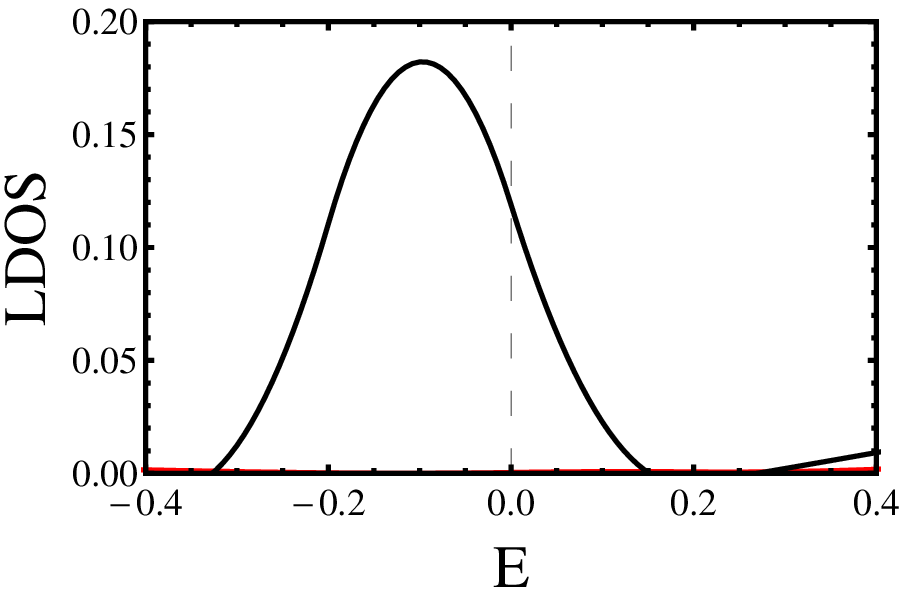}
\includegraphics[width=4.2cm,height=3.5cm]{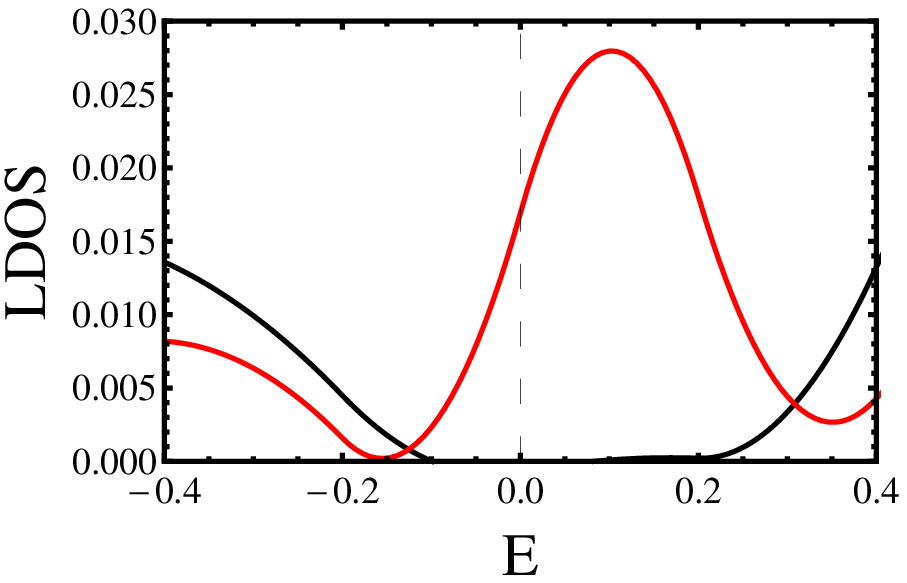}
\includegraphics[width=4.2cm,height=3.5cm]{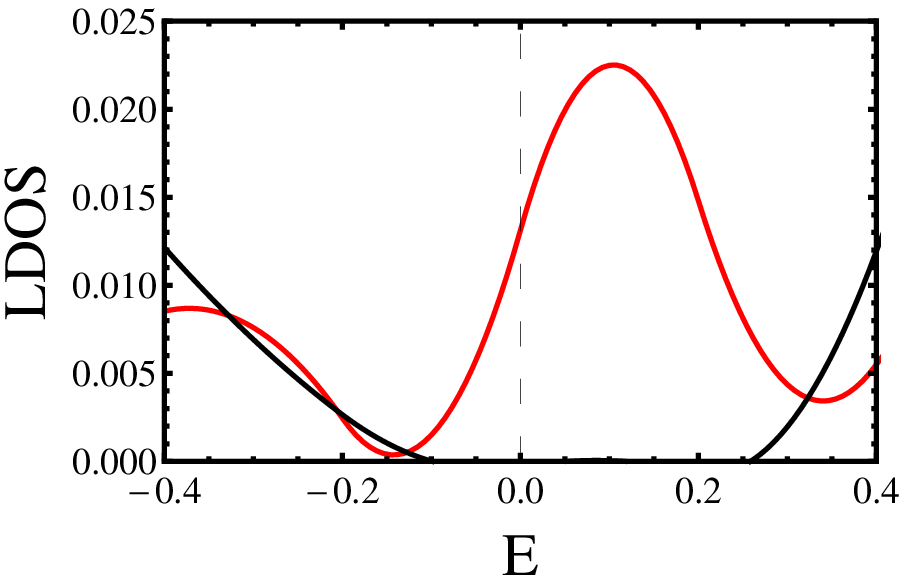}
\includegraphics[width=4.2cm,height=3.5cm]{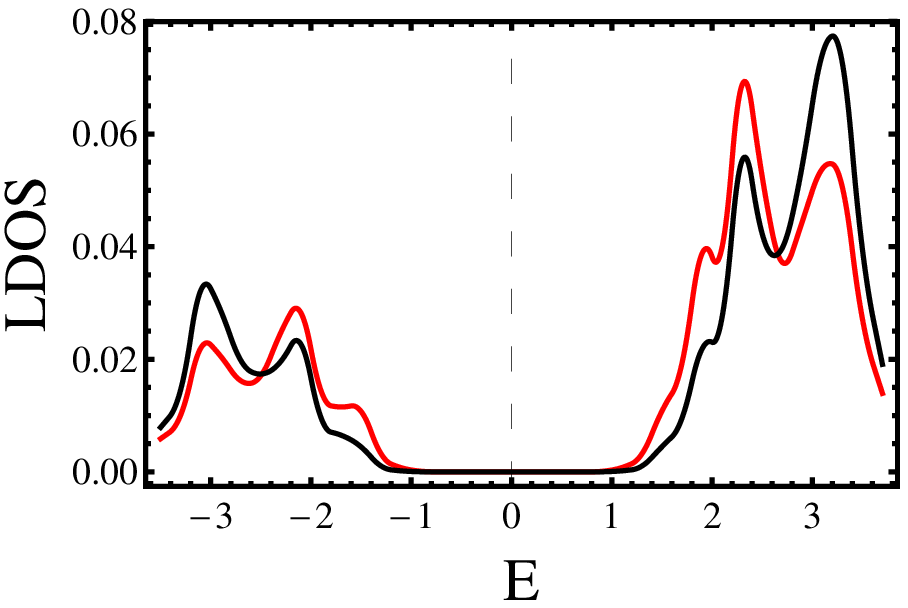}
\caption{(Color online) Plots of LDOS on A (red), and B(black) sublattices for $V_1=0.2$ and $q=0.02$. Choices of the location of sites on two sublattices in different plots are same as in Fig.~2. Here as well $E$ is measured in units of $t$, and $\Delta E=0.1$.}
\label{IntLDOS}
\end{figure}

\section{charge-density-wave ordering in strained graphene}

The existence of a large number of weakly dispersing modes near the zero energy in strained graphene can, in principle, trigger ordering tendencies for the Dirac fermions even when the interactions are weak. The $SU_c(2)$ chiral symmetry of the Dirac Hamiltonian $H[0]$ in the absence of any axial magnetic fields, generated by $\{ \gamma_3,\gamma_5, i\gamma_3 \gamma_5\}$, gets lowered to a $U_c(1)$ generated by $i \gamma_3 \gamma_5$, in its presence. On the other hand, $\left\{ H[\vec{a}], \gamma_0 \right\}=0$ dictates that the zero energy states have to be eigenstates of $\gamma_0$, and hence they select particular sublattice to live on. In addition, $H[\vec{a}]$ commutes with $\gamma_2 {\cal P}$, where $ {\cal P}p_2 \to -p_2$ and ${\cal P}a_2 \to -a_2$. Therefore, under $\gamma_2 {\cal P}$ the axial magnetic field changes its direction and the zero modes localized on two sublattices transform into each other. Notice that $\gamma_2(=\sigma_0 \otimes \sigma_1)$ exchanges two sublattices and zero energy manifold is composed of eigenstates of $\gamma_0$. Together these two symmetry generators ensures the \emph{sublattice degeneracy} of the zero energy subband, demonstrated in the last section\cite{lattice-zeromodes}.

Due to the inevitable doubling of the zero energy subband, arising from its support on two sublattices in the bulk and near the boundary, strained graphene may find itself in the CDW phase, when only half of these states, localized on one particular sublattice are filled, while those residing on the other sublattice are left empty. Thus CDW ordering in strained graphene also induces an imbalance of average electronic density between the bulk and boundary of the system. The CDW order parameter $\langle \Psi^\dagger \gamma_0 \Psi \rangle$, lacks the chiral symmetry, generated by $\gamma_3$ and $\gamma_5$, but preserves the time-reversal-symmetry, represented by $I_t=i\gamma_1 \gamma_5 K$, where $K$ is the complex conjugation\cite{HJR}. Therefore, in strained graphene formation of CDW order does not break any additional chiral symmetry, but \emph{spontaneously} lifts the sublattice degeneracy of the zero energy manifold, since $\left\{\gamma_0,\gamma_2 \right\}=0$. 
\begin{figure}[htb]
\includegraphics[width=8.5cm,height=5.5cm]{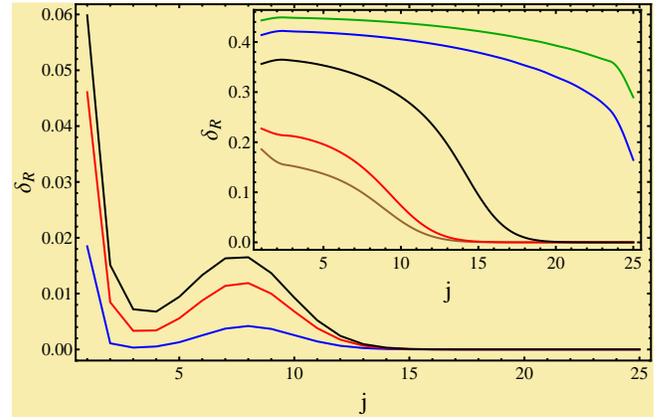}
\caption{ (Color online) Spatial variation of the \emph{local} CDW order $\delta_R$, defined in Eq.~(\ref{OPunitcell}), for $q=0.02$, and $V_1=0.2$(blue), $0.4$(red), $0.6$(black). Inset: Plot of the same quantity, but for $V_1=1.0$(brown), $1.25$(red), $1.5$(black), $2.0$(blue), $2.5$(green). Here $j$ is the row index of the cylinder shown in Fig.~1.}
\label{deltaRcylinder}
\end{figure}

Next through detailed numerical Hartree self-consistent calculation we establish that weak NN Coulomb repulsion drives the system into such an ordered phase. The interacting Hamiltonian with only the NN component of the Coulomb interaction is represented by 
\begin{equation}
H^{NN}_I= \frac{V_1}{2} \sum_{\langle k,l \rangle}\: \left( n_k -\frac{1}{2} \right) \; \left(n_l -\frac{1}{2}\right) -\mu N,
\end{equation}
where $n_k$ is the fermionic density at the $k^{th}$ site, $N$ is the total number of spinless fermions, $\mu$ is the chemical potential, and $\langle \cdots \rangle$ stands for the summation over the NN sites. Since $\left\{ H[\vec{a}],\gamma_0 \right\}=0$, the spectrum of $H[\vec{a}]$ is particle-hole symmetric and the charge-neutrality in the system is maintained by setting $\mu=0$. The usual Hartree decomposition leads to an effective single-particle Hamiltonian 
\begin{eqnarray}
H^{NN}_{SP}&=&V_1 \sum_{\langle k,l \rangle} \bigg[ \left( \langle n_{B,l}\rangle -\frac{1}{2} \right) \left(n_{A,k}-\frac{1}{2}\right) \nonumber \\
&+& \left( \langle n_{A,l}\rangle -\frac{1}{2} \right) \left(n_{B,k} -\frac{1}{2}\right)\bigg] -\mu N.
\end{eqnarray} 
Here $\langle n_{A(B)} \rangle$ is the average fermionic density at each site, and its \emph{local} deviation from the uniform background, can be captured by rewriting
\begin{equation}
\langle n_{A,k} \rangle=\frac{1}{2} + \delta_A(k) \quad \langle n_{B,k} \rangle=\frac{1}{2} -\delta_{B}(k).
\end{equation}
The overall charge-neutrality of the system $(\mu=0)$ then enforces the constraint 
\begin{equation}
\sum_{k} \delta_{A}(k)-\sum_{k} \delta_{B}(k)=0.
\end{equation}
We here keep $\delta_A$ and $\delta_B$ as functions of position and determine them self-consistently in a finite strained honeycomb lattice.

The Fock decompositions of the NN interaction, on the other hand, allow the local order parameters that couple two sublattices, such as translational symmetry breaking \emph{Kekule bond-density-wave order}\cite{hou-chamon-mudry}. However, due to the spatial separation of the zero modes, localized on two sublattices, such orders cannot develop any finite expectation value in strained graphene, at least when the interaction is weak. Henceforth, we completely neglect them for rest of the discussion and our Hartree self-consistent analysis for the CDW ordering with only NN component of the Coulomb repulsion serves the purpose of a \emph{variational calculation}.    
\begin{figure}[htb]
\includegraphics[width=4.2cm,height=3.5cm]{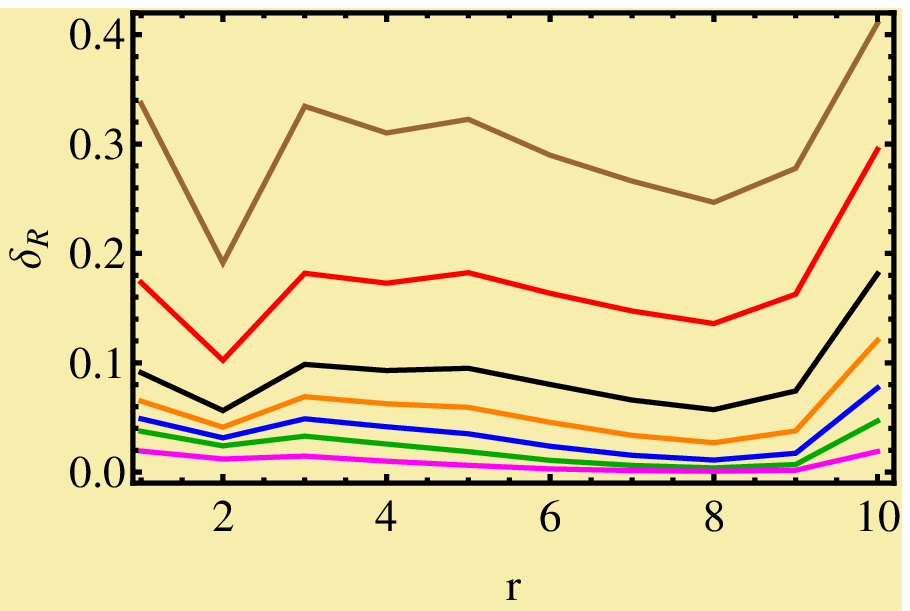}
\includegraphics[width=4.2cm,height=3.5cm]{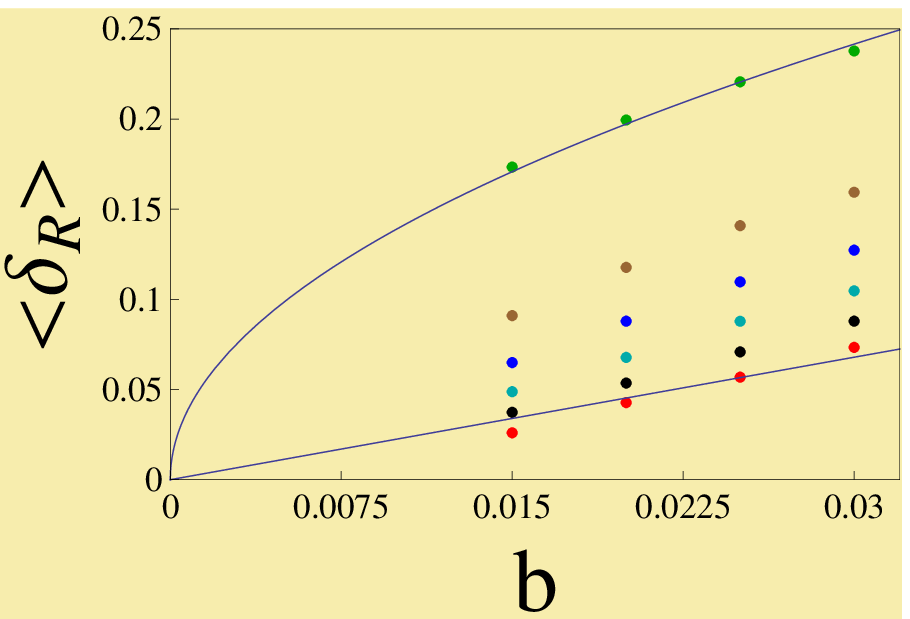}
\caption{(Color online) Left: spatial variation of $\delta_R(r)$ in a quasi-circular honeycomb lattice with open boundary for $q=0.0125$. Strength of the NN interaction reads as $V_1=0.2$, $0.3$, $0.4$, $0.5$, $0.6$, $0.76$, $0.9$, from bottom to top. Right: scaling of the average CDW order in the bulk of the system, $\langle \delta_R \rangle=\sum^{6}_{r=1} \delta_R(r)/6$, with the axial magnetic field ($b$). The strength of the NN interaction reads as $V_1=0.76$, $0.6$, $0.5$, $0.4$, $0.3$, $0.2$ from top to bottom. The bottom(top) solid line shows almost linear ($\sqrt{b}$) dependence of the gap with $b$ for $V_1=0.2(0.76=(V_1)_c)$. Here $b$ is measured in units of $b_0\sim 10^4$T, an axial field associated with the lattice spacing $c \sim 3 \mathring{A}$.}
\label{CDWopenboundary}
\end{figure}

Upon obtaining the self-consistent solutions of $\delta$s in a $36 \times 25$ cylindrical strained honeycomb lattice, we once again compute the LDOS on the sites of $A$ and $B$ sublattices at various locations of the cylinder. For $V_1= 0.2$ and $q=0.02$ results are shown in Fig.~3. Comparing Figs.~2 and ~3, we see that the peaks in the LDOS at zero energy in the non-interacting system get shifted to finite energies everywhere in the system when $V_1 \neq 0$. However, the shifts are of \emph{opposite} signs on two inequivalent sublattices. Therefore, maintaining the overall charge-neutrality of the system ($\mu=0$), but suppressing the density of states at the Fermi energy, a \emph{subcritical} NN interaction can place the system into an ordered phase by developing a density imbalance between the bulk and boundary of the system.

Next we demonstrate the spatial variation of the local CDW order parameter in a unit cell, defined as  
\begin{equation}\label{OPunitcell}
\delta_R(j)=\frac{1}{2} \bigg[ \delta_A(j)+\delta_B(j) \bigg], 
\end{equation}  
where $j$ is the row index, along the axis of the cylinder for $0.2 \leq V_1 \leq 2.5$, shown in Fig.~4. Therefore, for sufficiently weak interaction, $V_1 \ll (V_1)_c \approx 0.76$\cite{herbut-roy-pseudocat, franz-weeks, grushin-1}, $\delta_R$ develops nonzero expectation value only where the zero energy modes are localized, i.e. in the lower half of the cylinder ($j \leq j_{Max}/2$, with $j_{Max}=25$ in our system). Otherwise, for vanishingly small interaction the CDW order continuously disappears from the system. On the other hand, as NN interaction gets stronger ($V_1 \geq (V_1)_c$) the CDW ordering gradually starts to develop everywhere in the system, and finally for sufficiently strong NN repulsion ($V_1 \gg (V_1)_c$), the effect of the axial magnetic field becomes irrelevant, and a roughly uniform CDW order sets in everywhere in the system [see Fig.~4(inset)]. 
\begin{figure}[htb]
\includegraphics[width=4.2cm,height=3.5cm]{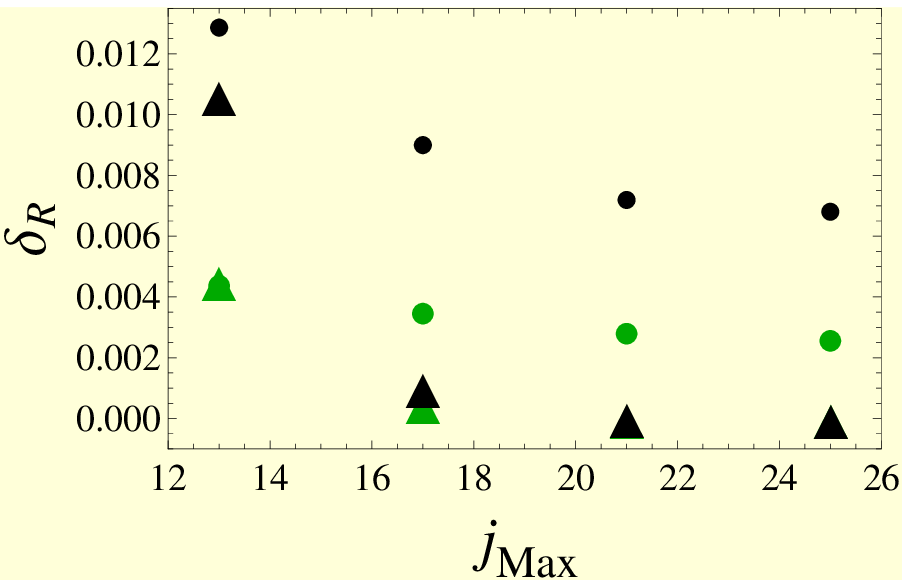}
\includegraphics[width=4.2cm,height=3.5cm]{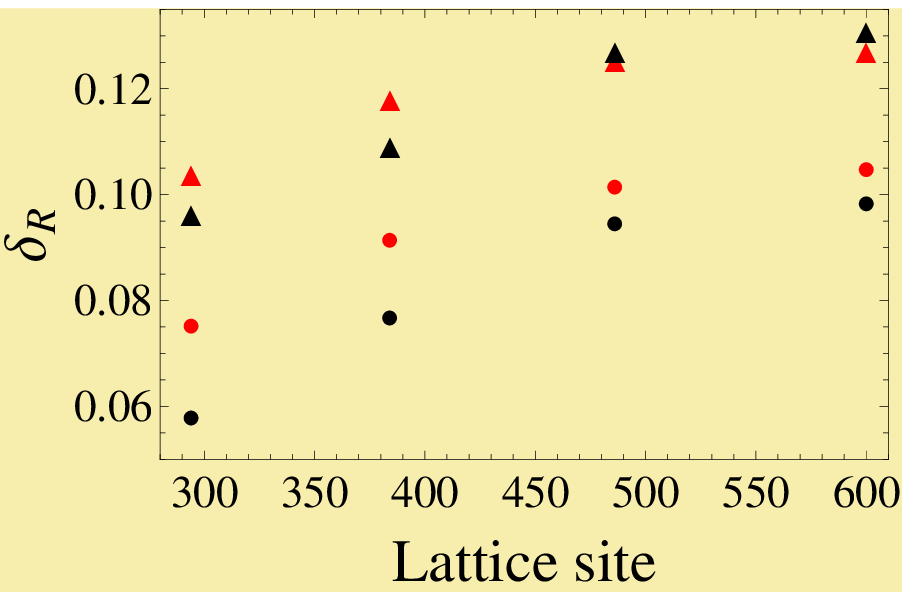}
\caption{(Color online) Left: Finite size scaling of the local CDW order parameter near the center of the cylinder (circles), at the top edge (triangles), for $V_1=0.2$(green), $0.4$(black), and $q=0.02$. Here $j_{max}$ corresponds to the maximal number of row in cylindrical graphene system. Right: finite size scaling of the same quantity in honeycomb lattice with open boundary, at the center (red), and middle (black) of the system for $V_1=0.3$ (circles), $0.4$ (triangles), and $q=0.015$.}
\label{finitesize}
\end{figure}

Moreover, the formation of CDW order in strained graphene does not depend on the geometry of the system. From the self-consistent Hartree solutions of $\delta$s, obtained in a \emph{quasi-circular} strained honeycomb lattice of 600 sites, with open boundary, we compute the CDW order in a unit cell of the $r^{th}$ quasi-circular ring ($\delta_R(r)$) around the center of the system. Results are shown in Fig.~5(left), for a wide range of the NN interaction, clearly showing that the CDW order can also be realized in finite strained graphene flake with open boundary for NN interactions, as weak as  $V_1=0.2 \ll (V_1)_c$. In this system, we implement the particular modulation of NN hopping amplitude, suggested in Ref.\onlinecite{herbut-roy-pseudocat} that introduces a finite axial magnetic field.

The average CDW order in the bulk of this system ($r \leq 6$) scales linearly with the axial magnetic field ($b$), for sufficiently weak NN repulsion ($V_1 \ll (V_1)_c$), which then reverts to a sublinear one for intermediate strength of the interaction. Finally an almost $\sqrt{b}$ scaling of the gap emerges for $V_1=(V_1)_c$, as shown in Fig.~5(right). Hence, the scaling of the CDW order with the axial magnetic field, is qualitatively similar to the one with the true magnetic fields\cite{herbut-roy-analysitc-scaling, roy-herbut-inhomogeneous}. Otherwise, the local CDW order at different region of the system, in the presence of axial magnetic fields becomes quite insensitive to the finite size effects, when the number of rows in cylindrical strained graphene ($j_{Max}$) is $\geq 20$, shown in Fig. 6 (left). Similar insensitivity to finite system size is reached when the total number of site is $\geq 484$, in the presence of open boundary, as shown in Fig. 6 (right).

\section{$V_1-V_2$ model in strained graphene}

In addition to CDW order, discussed in the previous section, resulting from the NN repulsion $(V_1)$, spinless fermions in strained graphene have also been predicted to support QAHI in the presence of weak NNN interaction $(V_2)$\cite{herbut-pseudocatalysis, thesis, ashwin, herbut-roy-pseudocat}. The QAHI order parameter reads as $\langle\Psi^\dagger i \gamma_1 \gamma_2 \Psi \rangle$, and it breaks the time-reversal symmetry, but represents a \emph{chiral-scalar} Dirac mass\cite{HJR}. In honeycomb lattice QAHI corresponds to an intra-sublattice circulating currents, orienting in opposite directions on two sublattices\cite{haldane}. Therefore spinless fermions in strained graphene can condense into two distinct ordered states, depending on the relative strength of $V_1$ and $V_2$. In the mean-field level, we next show that roughly when $V_1 >V_2$, the CDW ordering preempts the appearance of topological QAHI, when $\mu=0$. Nevertheless, by placing the chemical potential close to the first excited state (as $\mu \to \mu_1$) system can further lower its energy by developing the QAHI, shown in Fig.~7(a). If, on the other hand, the second neighbor repulsion dominates over the NN one, the topological QAHI can be realized at the Dirac point ($\mu=0$), whereas the CDW appears when the zero energy subspace is 3/4 or 1/4 filled (i.e. when $\mu \to \pm \mu_1$), as demonstrated through Fig.~7(c). 
\begin{figure}[htb]
\includegraphics[width=8.50cm,height=4.0cm]{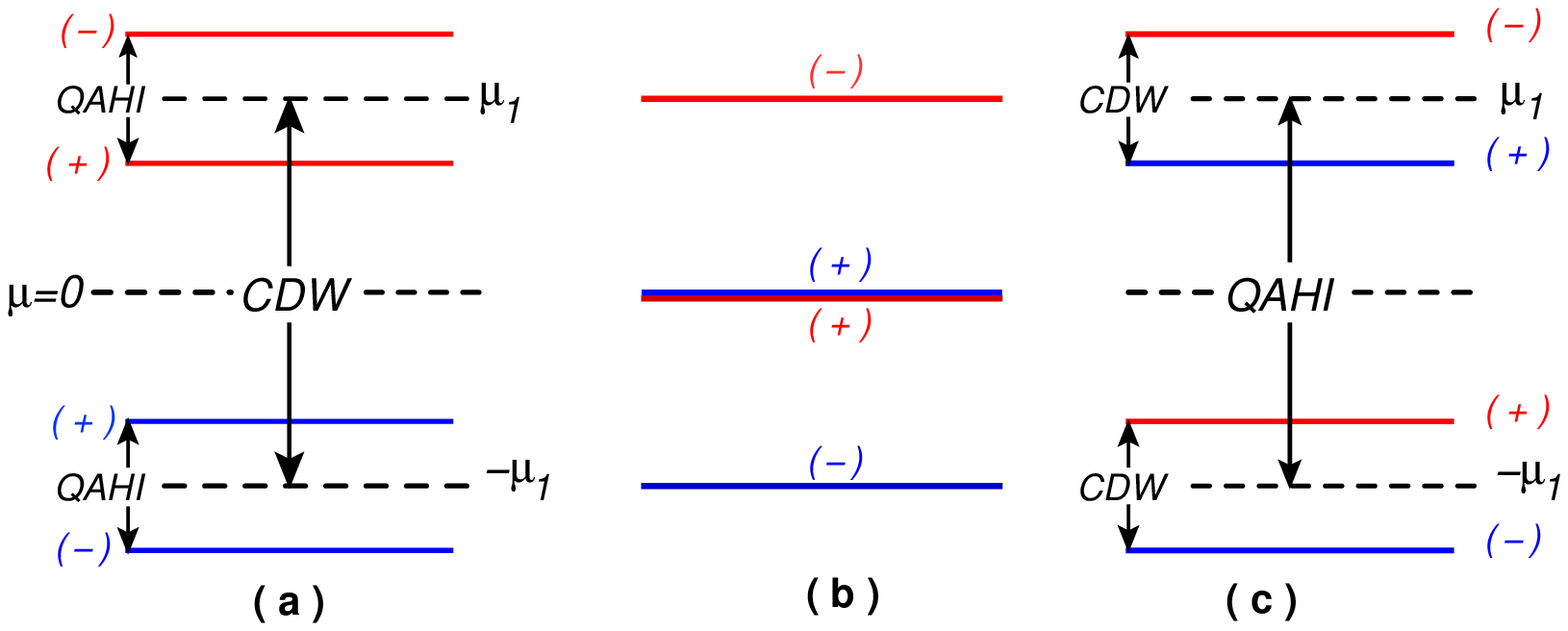}
\includegraphics[width=5.00cm,height=4.0cm]{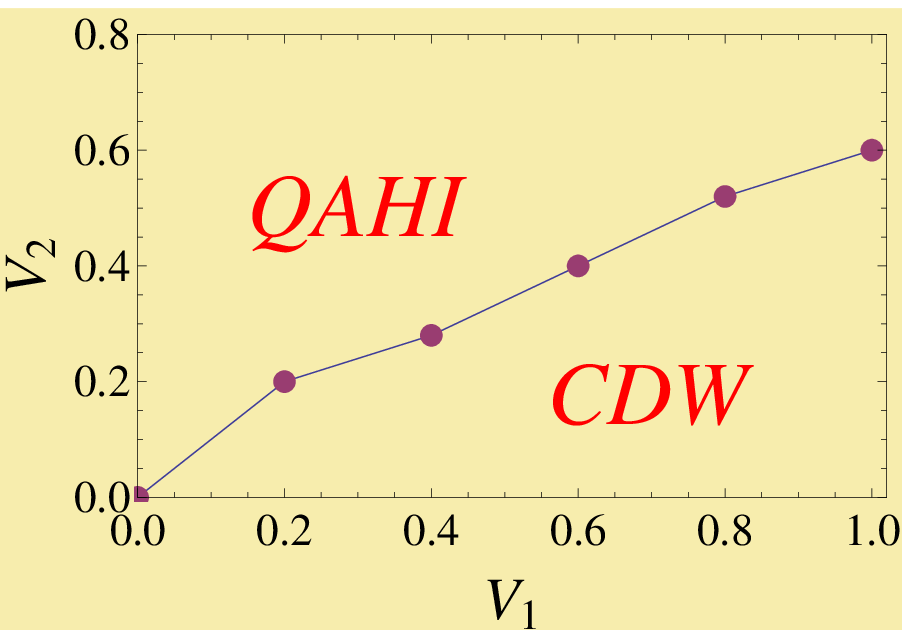}
\caption{ (Color online) Top: splitting of the zero energy states in strained graphene (for spinless fermions), when (a) $V_1 \gg V_2$, (b) $V_1 \sim V_2$, and (c) $V_2 \gg V_1$. States localized on A(B) sublattice are shown in blue(red), and $\pm$ correspond to their valley index $(\pm \vec{Q})$. Bottom: a representative mean-field phase diagram of $V_1-V_2$ model in strained graphene, when the chemical potential is pinned at the charge-neutrality-point.}
\label{V1V2}
\end{figure}

Next we address the competition between these two orderings for spinless fermions within the framework of a $V_1-V_2$ model in \emph{half-filled} deformed graphene. Keeping the chemical potential pinned at the Dirac point ($\mu=0$), if one gradually increases the strength of NNN repulsion, staring from the CDW phase, roughly around $V_1 \sim V_2$ a Landau level crossing takes place (see Fig.~7(b)), beyond which a QAHI sets in. It should be noted that when $\mu=0$ and CDW(QAHI) appears at the Dirac point, the splitting of the shifted zero modes due to the QAHI(CDW) around $\mu=\pm \mu_1$, shown in Fig.~7(a(b)), is absent. The NNN interaction among the fermions (spinless) can be represented by the interacting Hamiltonian
\begin{equation}\label{NNNHamil}
H^{NNN}_I=V_2 \sum_{\langle \langle k,l \rangle \rangle} n_k n_l -\mu N,
\end{equation}          
where $\langle \langle \cdots \rangle \rangle$ stands for the summation over the NNN  sites of the honeycomb lattice. Once again we set $\mu=0$ to maintain overall charge-neutrality of the system. The Fock decomposition of $H^{NNN}_I$ leads an effective single-particle Hamiltonian 
\begin{equation}
H^{NNN}_{SP}=\left( V_2 \sum_{\langle \langle k,l \rangle \rangle} \eta_{kl} c^\dagger_l c_k + H.c. \right)-\mu N, 
\end{equation} 
where $\eta_{kl}=\langle c^\dagger_k c_l \rangle$ is set to be \emph{imaginary}. Therefore, $\eta$ breaks the time-reversal-symmetry \cite{TRS-realspace} and corresponds to the QAHI, yielding Haldane's intra-sublattice circulating current. Recently it has been shown that when graphene flake is subject to axial magnetic fields sufficiently weak NNN interaction ($V_2 \ll (V_2)_c \approx 1.26$ \cite{franz-weeks, grushin-1}) can support QAHI. In the QAHI phase zero energy states living on A(B)-sublattice and localized near the $+(-)\vec{Q}$ valley are occupied while rest are left empty, for example \cite{herbut-roy-pseudocat}.

Recall that the CDW keeps all the zero energy states on A(B)-sublattice occupied(empty), and the matrices appearing in the definition of these two order parameters ($\gamma_0$ and $i \gamma_1 \gamma_2$) commute with each other. Thus these two orders for spinless fermions in a \emph{half-filled} strained graphene compete with each other, but \emph{they do not coexist}. Competition between the CDW and QAHI at half-filling can therefore be addressed by comparing the free energies per unit cell in the presence of these two orderings separately, which respectively read as 
\begin{equation}
\begin{split}
&\Delta F_{CDW}=\bigg[ -\sum^{n_{max}}_{n=1} \frac{|e^{CDW}_n|}{n_{max}}+ \sum^{{\cal R}}_{r=1} \frac{|\delta_R(r)|^2}{4 V_1 {\cal R}} \bigg]-F_N, \\ 
&\Delta F_{QAHI}=\bigg[-\sum^{n_{max}}_{n=1} \frac{|e^{QAHI}_n|}{n_{max}}+ \sum^{{\cal R}}_{r=1} \frac{|\eta(r)|^2}{4 V_2 {\cal R}}\bigg]-F_N.
\end{split} 
\end{equation} 
$F_N=-\sum^{n_{max}}_{n=1}|e_n|/n_{max}$ is the free energy per unit cell in the absence of any ordering. $n_{max}$ is half of the total number of sites ($\equiv$ number of unit cells) in the system and $(-e_n)$ represents the energy of the $n^{th}$ filled lavel. $-(e^{CDW}_n/e^{QAHI}_n)$ represents the same quantity, but in the presence of CDW/QAHI ordering. We here restrict ourselves with quasi-circular honeycomb lattice, with open boundary, where ${\cal R}$ represents the total number of quasi-circular rings in the system. $\delta_R(r)$, $\eta(r)$ are respectively the self-consistent CDW and QAHI order parameters per unit cell in the $r^{th}$ ring around the center. Comparing the free energies obtained from the self-consistent solutions of these two order parameters in a system with 600 sites (${\cal R}=10$), in the presence of roughly uniform axial magnetic field (with $q=0.015$), we extract a representative phase diagram in the $V_1-V_2$ plane, shown in Fig.~7 (bottom). Thus when the NN interaction substantially dominates over the NNN one, system finds itself in the CDW phase, and vice-versa. At the transition between these two states only one of the Dirac points remains fully gapped while the other one becomes gapless and system then simultaneously supports finite $\sigma_{xx}$ and quantized $\sigma_{xy}=e^2/2h$. Across the phase boundary between CDW and QAHI although the zeroth Landau levels go through a smooth \emph{level crossing} (see Fig. 7), both the order parameters jump discontinuously. Therefore, within the framework of \emph{restricted Hartree-Fock} analysis, we perform here, the transition between CDW and QAHI states is \emph{discontinuous} in nature.

It is worth mentioning that in pristine graphene CDW and QAHI are separated by either the semimetallic Dirac vacuum (when $V_1, V_2$ are weak) or the Kekule bond-density-wave order (when strong $V_1, V_2$ are comparable)\cite{franz-weeks, grushin-1}. Thus a direct transition between CDW and QAHI gets avoided in strain-free graphene. On the other hand, due the presence of a large number of states near the charge-neutrality point, Dirac semimetal gets vanquished in strained graphene for arbitrarily weak interactions, whereas the Kekule order cannot develop any finite expectation value at least for weak repulsions due to the spatial separation of the zero modes localized on two sublattices. Therefore, strained graphene offers an ideal platform to study a topological phase transition between QAHI and CDW at weak couplings, the exact nature of which, although believed to be a continuous one in the absence of axial fields \cite{gracey, witterich}, remains enigmatic thus far. Therefore, unrestricted Hartee-Fock analysis can provide further insight into the exact nature of this transition in strained graphene, which we leave for future investigation.

Recently it has been argued that axial gauge potentials, $a^3_\nu$ and $a^5_\nu$ in Eq.~(\ref{genaxial}), arise from translation symmetry breaking CDW orders \cite{ghaemi-ryu-gopalakrishnan, translationexplanation, ourCDW}. These non-abelian components of the axial gauge fields can, for example, be realized by introducing an appropriate \emph{super-lattice potential} induced by the substrate on which graphene is placed \cite{su2recent}. It should be noted that a particular Hartree decomposition of the NNN repulsion $H^{NNN}_I$ in Eq.~(\ref{NNNHamil}) yields local expectation values of $a^3_\nu$ and $a^5_\nu$. However, we strongly believe that NNN repulsion selects the QAHI phase over the modulated CDW orders, at least when it is weak $(V_2 \ll (V_2)_c)$, for the following reason. Since the QAHI represents a chiral symmetric but time-reversal-odd Dirac mass, it pushes down all the filled states below the chemical potential, besides splitting the zero subband. On the other hand, $a^3_j$ and $a^5_j$ components of the $SU(2)$ chiral/axial gauge field, commute with the Dirac Hamiltonian $H[0]$ and possibly gain energy only by splitting the zeroth Landau level, and hence likely to be energetically inferior to QAHI.

\begin{figure*}[htb]
\includegraphics[width=16.00cm,height=8.0cm]{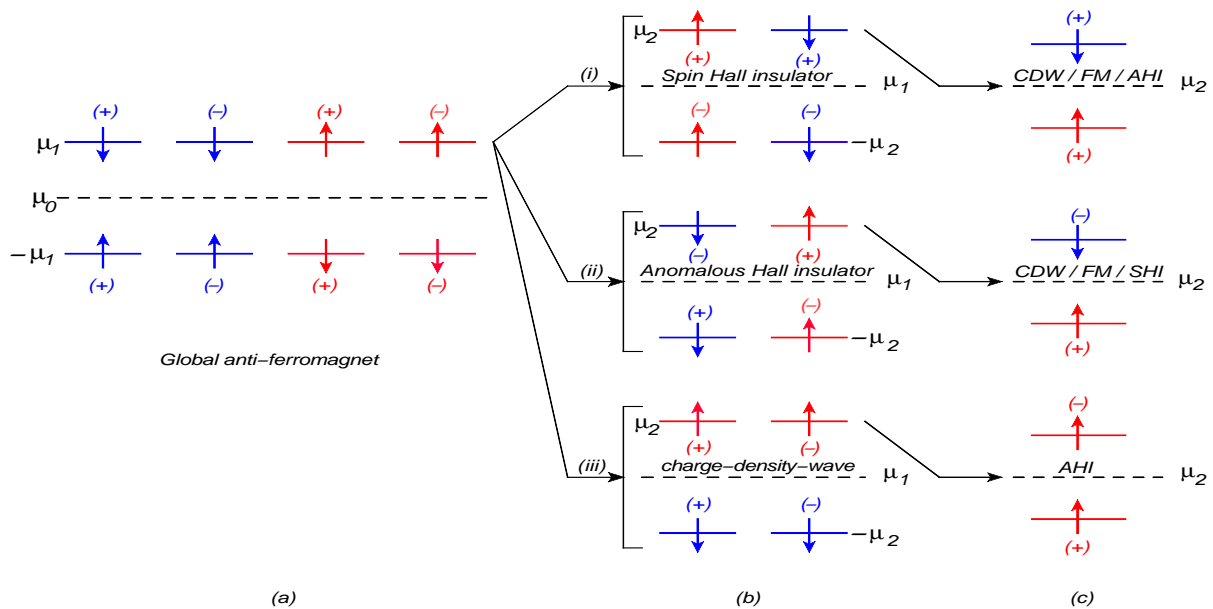}
\caption{ (Color online) Interaction driven splittings of the topological zero energy manifold in strained graphene. Here blue and red correspond to the zero energy states, localized on A and B sublattices, respectively, and $\pm$ to the ones localized near the Dirac valleys at $\pm \vec{Q}$. Projections of electron's spin along the z-axis are represented by $\uparrow, \downarrow$. (a) Leading instability at the Dirac point driven by global anti-ferromagnet order. (b) Subleading splittings driven by the ($i$) QSHI, ($ii$) QAHI and ($iii$) CDW, when the chemical potential $\mu \to \mu_1$. Splittings shown in panel (c) are followed by the ones shown in panel (b) and all the degeneracies of the zero energy subband is removed as $\mu \to \mu_2$.}
\label{axialsplit}
\end{figure*}

Through a detailed numerical analysis in pristine graphene, it has been argued that as one increases the strength of the NNN interaction\cite{grushin-1}, the Dirac semi-metal first enters into the QAHI phase when $V_2 \geq (V_2)_c$. Only upon further increasing $V_2 (\geq 2.3)$, system enters into the translational symmetry breaking CDW phase, where $\langle a^{3,5}_{1,2} \rangle \neq 0$. Therefore, variational approach, Monte-Carlo simulation with the NNN interaction in strained graphene can provide valuable insights into this competition. However, existence of a boundary (inevitably present in strained graphene) will possibly play a crucial role in these studies, since QAHI supports chiral edge states near the boundary. Interestingly, exact diagonalization studies in small clusters with periodic boundary have not found any sign of topological ordering in graphene \cite{grushin-2, daghofer}, while a similar study in the presence of an open boundary clearly exhibits the existence of QAHI state in pristine graphene \cite{tania}. Appearance of QAHI phase at strong enough $V_2$ in pristine graphene has also been supported in a variational Monte-Carlo simulation \cite{tania}.

\section{Spinful fermions: competing orderings in zero energy subband}

We now discuss the generic splitting of zeroth Landau level upon including all three, sublattice, valley and spin, degrees of freedom. Assuming that onsite Hubbard repulsion is the dominant component among the finite-range Coulomb repulsions, various possible splittings of the zero energy subband are shown in Fig. 8. With the restoration of fermion's spin additional orders that lack $SU(2)$ spin rotational symmetry can also set in at weak interactions. The possible orders include a magnetic ground state\cite{ashwin, roy-assaad-herbut}, and quantum spin Hall insulator(QSHI) \cite{abanin-pesin}. Although a Hund's flat band ferromagnet may appear as a natural ground state for Hubbard model in strained graphene, which, nevertheless, is true \emph{locally}, recent numerical simulations suggest that the magnetic ordering near the Dirac point contains a richer structure. In the magnetic phase, the local ferromagnetic order changes its sign as one approaches the boundary from the bulk, which then leads to net zero \emph{space-integrated} ferromagnetic moment. On the other hand, due to their support on opposite sublattices in these two regions, the zero energy states also give rise to an antiferromagnet order, which, on the other hand, is of same sign in the entire system, and the true magnetic ground state is an antiferromagnet, however, only \emph{globally}\cite{roy-assaad-herbut}. Such ordering possibly leads to dominant instability near the charge-neutrality point, since the onsite repulsion is the strongest component of the Coulomb interaction\cite{katsnelson}, and the resulting splitting of the pseudo zeroth Landau level is shown in Fig.~8(a).

When the NN repulsion is stronger than the NNN one, the dominant subleading splitting of the zeroth pseudo Landau level as the chemical potential $\mu \to \mu_1$, takes place through the formation of CDW, as shown in Fig.~8(b, iii). In addition, it also causes complete spin polarization of the remaining zero energy subband, and simultaneously supports a ferromagnet moment. Upon taking the chemical potential further away from the Dirac point, as $\mu \to \mu_2$ in Fig.~8(c, iii), QAHI is stabilized for arbitrarily weak NNN interaction.

A slightly longer range Coulomb interaction, the second nearest-neighbor repulsion ($V_2$) among the electrons can, in principle, support QSHI as well as QAHI; both leading to relativistic mass gaps in pristine graphene when $V_2$ is sufficiently strong\cite{raghu}. Otherwise, the orientation of the intra-sublattice circulating current in the \emph{time-reversal-symmetric} QSHI phase changes its sign upon exchanging the spin-projection as well as sublattice. However, due to the presence of aforementioned large density of states in the vicinity of the Dirac point, an interplay among these two orders can be realized even for weak $V_2$, when the graphene flake is subject to axial magnetic fields. The dominant subleading splittings of the zero subband driven by these two orders are shown in Figs.~8 (b, i) and 8 (b, ii), respectively. Although these two ground states are exactly degenerate at the mean-field level, upon analyzing the strength of the pertinent renormalized interaction couplings at a scale $l_b (\sim 1/\sqrt{b}) \gg c$, where $l_b$ is the axial magnetic length and $c$ is the lattice spacing, it has been argued that the tendency towards the formation of the QSHI is \emph{slightly} favored over the QAHI\cite{herbut-roy-pseudocat}, similar to the situation at strong couplings in strain-free graphene\cite{raghu}. Thus, at least when $V_2>V_1$, and $\mu \to \mu_1$, system can further lower its energy by developing QSHI, which supports a \emph{quantized spin Hall conductance} $2e^2/h$. Finally, upon taking the chemical potential close to the QSHI/QAHI gap ($\mu \to \mu_2$, see Fig.~8 (c, i/ii)), a CDW order develops in the system, which once again fully polarizes the spin of remaining zero modes. Otherwise, depending on the appearance of QSHI or QAHI as $\mu \to \mu_1$, CDW realized at $\mu=\mu_2$ supports a quantized ($e^2/h$) charge or spin Hall conductance, due to the presence of an accompanying anomalous or spin Hall insulator order, respectively. Therefore, depending on the relative strength of various finite range components of the Coulomb interaction, which possibly can be tuned to certain degree at least in \emph{molecular graphene}\cite{manoharan-polini}, strained graphene may exhibit a wide variety of interaction driven splitting of the topological zero energy subband.

\section{Summary and discussion}

To summarize, we here demonstrate that strain in graphene, through binding a large number of states at the Dirac point, can be conducive for various orderings, in the presence of weak repulsive interactions. These special zero energy modes, however, reside on opposite sublattices in the bulk and near the edge of the system. A detailed numerical Hartree self-consistent calculation shows that weak NN interaction ($V_1$) can remove the sublattice degeneracy from the zero energy subband by spontaneously developing a density imbalance between the bulk and boundary, which simultaneously generates a staggered pattern of charge among two sublattices of honeycomb lattice as well. The CDW order exhibits a crossover from linear to sublinear scaling with the axial field $(b)$ as the subcritical NN repulsion gets stronger and an almost perfect $\sqrt{b}$ scaling emerges at zero axial field criticality $(V_1=(V_1)_c)$, resembling in this regard its scaling in real magnetic field\cite{herbut-roy-analysitc-scaling, roy-herbut-inhomogeneous}, and the scaling of QAHI in axial fields\cite{herbut-roy-pseudocat}.

Taking the valley and spin degrees of freedom into account, we argue that various other orders, such as a magnetic ground state, QAHI, QSHI, can also set in at and near the Dirac points in strained graphene. While the onsite repulsion supports a magnetic ground state, the NNN interaction ($V_2$) can stabilize QAHI or QSHI phase. Although various works studied the instability of Dirac fermions in strained graphene previously in the presence of NNN and onsite repulsion, respectively supporting topological insulators \cite{herbut-pseudocatalysis, thesis, ashwin, herbut-roy-pseudocat} and magnetic order \cite{ashwin, roy-assaad-herbut}, competitions among these orders as well as that with the CDW order have been overlooked. In this work we address these issues in details. Within the framework of a $V_1-V_2$ model for the spinless fermions, we here address the competition between CDW and QAHI in a half-filled strained honeycomb lattice and show that roughly when $V_2 > V_1$ the QAHI is energetically favored over CDW and vice-versa. Possibly the onsite repulsion is the strongest Coulomb interaction and drives the zero energy subband in strained graphene towards the formation of an unconventional magnetic ordering, \emph{global anti-ferromagnet}, at the Dirac point. Nevertheless, the remaining orders, namely CDW, QAHI and QSHI, lead to further splitting of the shifted zero modes, upon continuously tuning the chemical potential away from the charge-neutrality point.

Although our discussions strictly focuses on the effect of time-reversal-symmetric axial magnetic fields, they are equally applicable even when a true (time-reversal-symmetry breaking) magnetic field penetrates strained graphene, but the axial one is stronger. The Dirac Hamiltonian in Eq.~(\ref{HDpseudo}), in the presence of both real and pseudo magnetic field generalizes to\cite{roy-oddQHE}  
\begin{equation}
H[\vec{a},\vec{A}]=\sigma_0 \otimes i\gamma_0 \sum_{\nu=1,2} \gamma_\nu \left(\hat{p}_\nu -A_\nu-i \gamma_{3} \gamma_{5} a_\nu (\vec{x}) \right),            
\end{equation}
upon restoring the spin degrees of freedom (represented by the first Pauli matrix $\sigma_0$). The real magnetic field $B$ is given by $\vec{B}=\vec{\nabla} \times \vec{A}$, and it is set to be perpendicular to graphene plane, thus $\vec{B}=B \hat{z}$. When both the fields are uniform and $b>B$, the effective magnetic fields at two the valleys ($\pm \vec{Q}$) $(b \pm B)$, still point in opposite directions, and the Landau levels at finite energies appear at $\pm E_{n,\pm}$ for each spin component, where $E_{n,\pm}=\sqrt{2 n (b \pm B)}$, with $n=1,2,\cdots$. Even in this set up, system continues to host topologically protected zero energy subband $(n=0)$, and the areal degeneracy of all the Landau levels, localized near $\pm \vec{Q}$ valley is $(b \pm B)/ \pi$\cite{roy-hu-yang}. The zero energy states on $A (B)$ sublattice live in the bulk (near the boundary) of the system, as long as $b>B$. Therefore, onsite repulsion in this system as well can support the unconventional magnetic ground state at half-filling, shown in Fig.~8(a). Furthermore, the proposed interplay among CDW, QAHI, and QSHI orders, resulting in the subleading splittings of the shifted zero modes at finite doping, shown through Fig.~8(b) and 8(c), holds even when strained graphene is placed in a weak true magnetic field ($b>B$). However, due to the single-particle Zeeman coupling, represented by $H_Z=g B \left( \sigma_3 \otimes I_4 \right)$, where $g \approx 2$ in graphene, spin-rotational symmetry breaking orders, e.g. global antiferromagnet and QSHI, get projected onto the \emph{spin easy-plane}.

Therefore, strained graphene, apart from its potential applicability in semiconductor industry due to the resulting valley dependent gauge fields, and possibility of isolating the ``valley-charge" \cite{beenakker}, also provide a unique opportunity to observe various broken symmetry phases\cite{herbut-pseudocatalysis, herbut-roy-pseudocat, ashwin, roy-assaad-herbut, abanin-pesin}, as well as proximity induced unconventional superconductivity\cite{roy-juricic}. Recent progress in achieving strained induced axial magnetic fields in regular\cite{pseudofield-experiment-1, pseudofield-experiment-2} and molecular graphene\cite{artificialgraphene}, over a range $50-600$T, proposal of valley-gauge fields in cold atom systems\cite{opticallattice-axial}, possible connection of strained graphene with Dirac equation in curved space\cite{bo-yang}, together with the theoretical interests in understanding the competition among distinct topological ground states, their microscopic origins in low dimensional systems\cite{kane-mele-hubbard-1, kane-mele-hubbard-2, rachel, Lin-he}, make our study interesting, timely and experimentally relevant.

\acknowledgements

This work is supported by US-ONR and by LPS-CMTC. B. R. is thankful to Vladimir Juri\v ci\' c for critical reading of the manuscript.


\begin{thebibliography}{99}
\bibitem{v1v2units} Throughout this paper, $V_1$, $V_2$ are measured in units of $t(=1)$.
\bibitem{cylinder} In cylindrical graphene the top and the bottom edges are constituted by $A$ and $B$ sublattices, respectively, see Fig.~1. Therefore, two edges of the cylinder are inequivalent.
\bibitem{TRS-realspace} The time-reversal operator in real space is simply $K$ (complex conjugation).
\bibitem{ourCDW} Charge-density-wave order we are considering in this work preserves the translational symmetry, since $\left[ \gamma_0, i \gamma_3 \gamma_5 \right]=0$.
\end{thebibliography}
\end{document}